\documentclass[twocolumn,showpacs]{revtex4} 
\usepackage{amsmath}
\usepackage{amssymb}
\usepackage{graphicx,color}

\def\diff{\mathrm d}
\newcommand{\etal}{{\it et al.}\ }

\begin{document}

\title{
Decomposition of modified Landau-Lifshitz-Gilbert equation\\
and corresponding analytic solutions
}
\author{Taichi Kosugi}

\affiliation{Nanosystem Research Institute ``RICS", AIST, Umezono, Tsukuba 305-8568, Japan}

\begin{abstract}
The Suzuki-Trotter decomposition in general allows one to divide the equation of motion of a dynamical system
into smaller parts whose integration are easier than the original equation.
In this study, we first rewrite by employing feasible approximations the modified Landau-Lifshitz-Gilbert equation for localized spins in a suitable form
for simulations using the Suzuki-Trotter decomposition.
Next we decompose the equation into parts and demonstrate that the parts are classified into three groups,
each of which can be solved exactly.
Since the modified Landau-Lifshitz-Gilbert equation from which we start is in rather a general form,
simulations of spin dynamics in various systems accompanying only small numerical errors are possible.
\end{abstract}


\maketitle

\section{Introduction}

The recent development of fabrication techniques of nanostructures and the extension of knowledge about magnetic properties of condensed matters
are accelerating the efforts for achieving more efficient electric control of magnetic states for applications in engineering.
The Landau-Lifshitz-Gilbert (LLG) equation\cite{bib:2268_2,bib:2268_3} has been widely used for interpretation of experiments and theoretical analyses for complex magnetization dynamics.
In particular, its phenomenologically modified versions\cite{bib:2058,bib:2286,bib:2324} proposed for incorporating the effects of spin torque induced by an electric current\cite{bib:2058_3}
are important both for realization of spintronics devices and for micromagnetic simulations of such nonequilibrium phenomena.
Theoretical studies, including numerical simulations, often address the space in which magnetic moments reside as a continuum
since the typical length scale of magnetization patterns is on the order of ten nanometers, much larger than lattice constants.

On the other hand, there are many systems whose spin configurations are known to be crucially influenced by the microscopic arrangement of each spin
such as noncollinear spin systems, which include helical magnetism, weak ferromagnetism, and frustrated spins.
The realization of spin configuration in such systems often originate from the symmetry of crystal lattice
and some of them are related to relativistic effects on electrons, which are in general effective only in the vicinity of each ion,
demanding microscopic considerations.

For construction of a reliable model for a specific electronic system,
schemes for determination of its parameters using first-principles electronic structure calculations have been developed.
They allows one to calculate the effective exchange integrals\cite{bib:2181}
and the Dzyaloshinskii-Moriya interaction\cite{bib:DMinteraction} between localized spins.
Schemes taking the electronic correlation effects into account for calculations of the Coulomb repulsion and the Coulomb exchange integrals\cite{bib:downfolding}
between localized electronic orbitals have also been developed.
Even for spin simulations of large length scales, those which are the most reliable should be performed
using model parameters determined by first-principle calculations.
To perform simulations of spin dynamics by incorporating the effects coming from the microscopic arrangement of spins
and evaluate quantitatively the properties of the system,
not only the material-specific parameters but also reliable algorithms for numerical integration are needed.
One of the most reliable integration scheme is the Suzuki-Trotter decomposition (STD) method\cite{bib:2271},
whose numerical error is third-order with respect to the time step.
If we know the exact solutions of the individual parts of a decomposed time-development operator,
the STD method ensures that the source of numerical errors is only the STD method itself.

This paper is organized as follows.
In Sec. II,
we rewrite the modified LLG equation into a physically appropriate form for localized spins,
which is demonstrated to consist of terms of three kinds.
In Sec. III,
the analytic solution of the equation of motion for each of the parts is provided.

\section{Modified Landau-Lifshitz-Gilbert equation}

The LLG equation and its modified versions are for magnetization $\boldsymbol{M}$, which is defined as the total magnetic moment density
as a function of continuous space coordinate.
In the present study, we use the effective spin angular momentum $\boldsymbol{S} \equiv -v \boldsymbol{M}/\gamma$, where $v$ is an appropriate volume unit, rather than $\boldsymbol{M}$.
$v$ depends on the spatial scale for physical phenomena one would like to see and study.
We assume that the effective spin behaves as a single classical spin.
Although the validity of the choice of $v$ and the treatment of the single spin should be of course confirmed for individual simulations of specific systems,
such problems are out of scope of the present study.

In this subsection, we rewrite the phenomenologically modified LLG equation
by replacing the spatial derivatives with finite differences of spin variables for two reasons.
The first is that we are treating the system as consisting of localized spins which are not necessarily arranged on a regular mesh.
The second is that the rewritten parts of the equation of motion are of the form whose exact solutions are known.

\subsection{Original form}

We express the effective external field felt by the $n$-th spin in the system as
$
\boldsymbol{H}_n 
= \gamma \boldsymbol{B}_n
-\sum_{m \ne n} \mathcal{J}_{nm} \boldsymbol{S}_m
$.
$\gamma$ is the gyromagnetic ratio and $\boldsymbol{B}_n$ is the external magnetic field.
$\mathcal{J}_{nm}$ is a $3 \times 3$ matrix representing the effective exchange interaction between the $n$-th and $m$-th spins.
$m$ in the summation runs over all the spins surrounding the $n$-th spin.
By using a symmetric $3 \times 3$ matrix $\mathcal{J}_{n}^\mathrm{S}$,
we express the field due to the magnetocrystalline (single-ion) anisotropy as $-2 \mathcal{J}_n^{\mathrm{S}} \boldsymbol{S}_n$,
whose factor $2$ comes from the quadratic form of the spin Hamiltonian.\cite{bib:2315,bib:2312}
We introduce constraints on the magnitudes of the spins in the present study as
$S_n = \mathrm{const}$.
Thus $\mathcal{J}_{n}^\mathrm{S}$ can be assumed to be traceless without loss of generality.

Kim \etal\cite{bib:2324} recently derived a general form of the LLG equation for studying
magnetization dynamics in a thin ferromagnet incorporating the Rashba effect\cite{bib:2324_19},
which is a kind of spin-orbit interaction due to lack of inversion symmetry.
Meanwhile, Zhang and Zhang\cite{bib:2058} had proposed an enhanced damping tensor,
which gives rise to temporarily and spatially varying damping term, as a modification to the LLG equation.
It was introduced to describe the effect of a conducting electron in a ferromagnet which carries away the excess angular momentum of the precessing ferromagnetic moment.\cite{bib:2058_8,bib:2058_9}
The current-driven rotational motion of a domain wall is influenced significantly by the damping enhancement.\cite{bib:2347}

We adopt the following form of the modified LLG equation in the most general form to date as a starting point of the present work:
\begin{gather}
	\frac{\diff \boldsymbol{S}_n}{\diff t} 
	= \boldsymbol{S}_n \times 
		\Bigg( \boldsymbol{H}_n + \boldsymbol{H}_{\mathrm{R}} + \frac{\beta}{S_n} \boldsymbol{S}_n \times \boldsymbol{H}_{\mathrm{R}} \Bigg)
	\nonumber \\
		-2 \boldsymbol{S}_n \times \mathcal{J}_{n}^\mathrm{S} \boldsymbol{S}_n
		+ (\boldsymbol{u} \cdot \nabla) \boldsymbol{S}_n
		+ \frac{\beta}{S_n} \boldsymbol{S}_n \times (\boldsymbol{u} \cdot \nabla) \boldsymbol{S}_n
	\nonumber \\
		- \frac{1}{S_n} \boldsymbol{S}_n \times \mathcal{D} \frac{\diff \boldsymbol{S}_n}{\diff t}
	\label{LLGeq_1}
	,
\end{gather}
where 
\begin{gather}
	\mathcal{D}_{ij} = \alpha \delta_{ij} + \frac{\eta}{S_n^4} \sum_k
		( \boldsymbol{S}_n \times \partial_k \boldsymbol{S}_n )_i
		( \boldsymbol{S}_n \times \partial_k \boldsymbol{S}_n )_j
\end{gather}
$(i, j, k = x, y, z)$
is the enhanced damping tensor\cite{bib:2058} 
with the dimensionless Gilbert damping constant $\alpha$
and the parameter $\eta$ in a dimension of squared length.
The electric current density $\boldsymbol{j}_e$ and its polarization rate $P$ defines the velocity vector
$\boldsymbol{u} = \mu_\mathrm{B} P/[ M_s (1 + \beta^2)] \boldsymbol{j}_e$,
where $\beta$ is the ratio of the nonadiabatic spin transfer torque to the adiabatic one\cite{bib:2286,bib:2325}
and $M_s$ is the saturation magnetization.
$\mu_\mathrm{B}$ is the Bohr magneton.
$\boldsymbol{H}_{\mathrm{R}} = -k_{\mathrm{R}} (\boldsymbol{e}_z \times \boldsymbol{u} )$
is the effective field introduced for the Rashba effect,\cite{bib:2324}
whose strength is measured by $k_{\mathrm{R}}$ in a dimension of inverse length.

Since we have adopted the expression of the effective field as the summation of the contributions from the surrounding effective spins,
our formulation allows for an arbitrary mesh of space for coarse-graining.
Furthermore, the derivation of the equation of motion below is applicable to microscopic (not effective) spins distributed in a crystal.

\subsection{Rewritten form}

As stated above, the spatial derivatives in Eq. (\ref{LLGeq_1}) should be replaced with finite differences appropriate for specific systems.
One might simply calculate the spatial derivative of a spin direction by using its neighboring ones as
$
\partial_i \boldsymbol{S}_n \approx
\Delta_i^{n} \boldsymbol{S}_n
+ \sum_{m \ne n} \Delta_i^{nm} \boldsymbol{S}_m
\equiv
\hat{\Delta}_i \boldsymbol{S}_n
$.
The appropriate definitions of the vectors $\boldsymbol{\Delta}^n$ and $\boldsymbol{\Delta}^{nm}$
which characterize the numerical derivative operator $\hat{\Delta}_i$
depend on the locations of the spins in the system, including the dimensionality and the periodicity.
The constant-magnitude condition of the spin requires
that the spin and its spatial derivative be orthogonal since $\boldsymbol{S}_n \cdot \partial_i \boldsymbol{S}_n = \partial_i S_n^2 /2 = 0$.
The numerical derivative proposed above, however, does not ensure the orthogonality.
We therefore subtract the component parallel to $\boldsymbol{S}_n$ from $\hat{\Delta}_i \boldsymbol{S}_n$ as
$\hat{\Delta}_i^\perp \boldsymbol{S}_n \equiv \hat{\Delta}_i \boldsymbol{S}_n - ( \boldsymbol{S}_n \cdot \hat{\Delta}_i \boldsymbol{S}_n ) \boldsymbol{S}_n / S_n^2$.
We can thus write
\begin{gather}
	\hat{\Delta}_i^\perp \boldsymbol{S}_n
	= \hat{\widetilde{\Delta}}_i \boldsymbol{S}_n
		- \hat{\Delta}_i^\parallel \boldsymbol{S}_n
	,
	\label{nderiv_2}
\end{gather}
where
\begin{gather}
	\hat{\widetilde{\Delta}}_i \boldsymbol{S}_n
		\equiv \sum_{m \ne n} \Delta_i^{nm} \boldsymbol{S}_m
	,
	\\
	\hat{\Delta}_i^\parallel \boldsymbol{S}_n
		\equiv \sum_{m \ne n} \Delta_i^{nm} \frac{\boldsymbol{S}_n \cdot \boldsymbol{S}_m}{S_n^2} \boldsymbol{S}_n
		\equiv \Delta_{i n}^\parallel \boldsymbol{S}_n
	.
	\label{hat_para_Delta}
\end{gather}
It is noted that $\hat{\widetilde{\Delta}}_i \boldsymbol{S}_n$ does not contain the variable $\boldsymbol{S}_n$.
We adopt the constant-magnitude numerical derivative $\hat{\Delta}_i^\perp \boldsymbol{S}_n$ instead of $\partial_i \boldsymbol{S}_n$ in the present study.

Let us then rewrite the equation of motion by employing the constant-magnitude numerical derivative.
There clearly exists a relation
$\hat{\Delta}_i^\perp \boldsymbol{S}_n 	= -\boldsymbol{S}_n \times ( \boldsymbol{S}_n \times \hat{\widetilde{\Delta}}_i \boldsymbol{S}_n )/ S_n^2$,
which can be used for rewriting the spin transfer torque terms in Eq. (\ref{LLGeq_1}) as
\begin{gather}
	(\boldsymbol{u} \cdot \nabla) \boldsymbol{S}_n
		+ \frac{\beta}{S_n} \boldsymbol{S}_n \times (\boldsymbol{u} \cdot \nabla) \boldsymbol{S}_n
	\nonumber \\
	=
	\beta \boldsymbol{S}_n \times \widetilde{\boldsymbol{H}}_n
	-\frac{1}{S_n} \boldsymbol{S}_n \times ( \boldsymbol{S}_n \times \widetilde{\boldsymbol{H}}_n )
	,
\end{gather}
where we have defined
\begin{gather}
	 \boldsymbol{\widetilde{H}}_n \equiv
		\frac{1}{S_n} \sum_i u_i \hat{\widetilde{\Delta}}_i \boldsymbol{S}_n
		= \frac{1}{S_n} \sum_{m \ne n} ( \boldsymbol{u} \cdot \boldsymbol{\Delta}^{nm} ) \boldsymbol{S}_m
	.
\end{gather}

The equation of motion of the form Eq. (\ref{LLGeq_1}) is not tractable since the time derivative appears on both sides.
To make the equation computationally manageable, we operate $\boldsymbol{S}_n \times \mathcal{D}$ on both sides of Eq. (\ref{LLGeq_1}).
We calculate the right hand side below by dividing it into four parts.

The first and second parts are calculated as follows.
We neglect the terms which involves $\eta H_\mathrm{R}$ and $\eta \mathcal{J}^\mathrm{S}$
since they originate from the spin-orbit interactions and their energy scale is much smaller than those for nonrelativistic interactions in general
and the effects of these terms on the qualitative behavior of enhanced damping are expected to be insignificant.
Thus we can rewrite the first part as
\begin{gather}
	\boldsymbol{S}_n \times \mathcal{D} \Bigg[ \boldsymbol{S}_n \times
		\Bigg( \boldsymbol{H}_n + \boldsymbol{H}_{\mathrm{R}} + \frac{\beta}{S_n} \boldsymbol{S}_n \times \boldsymbol{H}_{\mathrm{R}} \Bigg)
		\Bigg]
	\nonumber \\
	= \alpha \boldsymbol{S}_n \times [ \boldsymbol{S}_n \times ( \boldsymbol{H}_n + \boldsymbol{H}_{\mathrm{R}} ) ]
	-	\alpha \beta S_n \boldsymbol{S}_n \times \boldsymbol{H}_{\mathrm{R}}
	\nonumber \\
		+ \frac{\eta}{S_n^2} \sum_i
		( \hat{\Delta}^\perp_i \boldsymbol{S}_n \cdot  \boldsymbol{H}_n )
		\boldsymbol{S}_{n} \times
		( \boldsymbol{S}_n \times \hat{\widetilde{\Delta}}_i \boldsymbol{S}_n )
	\label{SD_part_1}
\end{gather}
and the second part as
$-2 \boldsymbol{S}_n \times \mathcal{D} ( \boldsymbol{S}_n \times \mathcal{J}^\mathrm{S}_n \boldsymbol{S}_n )
= -2 \alpha  \boldsymbol{S}_n \times ( \boldsymbol{S}_n \times \mathcal{J}^\mathrm{S}_n \boldsymbol{S}_n )$.
Furthermore, we neglect terms which involves $\beta \eta$ for the third part since $\beta$ is much smaller than unity\cite{bib:2325},
and it is rewritten as
\begin{gather}
	\boldsymbol{S}_n \times \mathcal{D} (\boldsymbol{u} \cdot \nabla) \boldsymbol{S}_n
		+ \frac{\beta}{S_n} \boldsymbol{S}_n \times \mathcal{D}
		[
			\boldsymbol{S}_n \times (\boldsymbol{u} \cdot \nabla) \boldsymbol{S}_n
		]
	\nonumber \\
	= 
		\alpha S_n \boldsymbol{S}_n \times \widetilde{\boldsymbol{H}}_n
		+ \alpha \beta \boldsymbol{S}_n \times ( \boldsymbol{S}_n \times \widetilde{\boldsymbol{H}}_n )
	\nonumber \\
		+ \frac{\eta}{S_n^3} \sum_i
			\boldsymbol{S}_n \cdot ( \hat{\widetilde{\Delta}}_i \boldsymbol{S}_n \times \widetilde{\boldsymbol{H}}_n )
			\boldsymbol{S}_n \times ( \boldsymbol{S}_n \times \hat{\widetilde{\Delta}}_i \boldsymbol{S}_n )
		.
	\label{SD_part_3}
\end{gather}
The fourth part to the first order of $\eta$ is rewritten as
\begin{gather}
	-\frac{1}{S_n} \boldsymbol{S}_n \times \mathcal{D} \Bigg( \boldsymbol{S}_n \times \mathcal{D} \frac{\diff \boldsymbol{S}_n}{\diff t} \Bigg)
	\nonumber \\
	=
		\alpha^2 S_n \frac{\diff \boldsymbol{S}_n}{\diff t}
		+ \frac{\alpha \eta}{S_n^3} \sum_i
		[ 
		S_n^2 ( \hat{\Delta}^\perp_i \boldsymbol{S}_n ) ( \hat{\Delta}^\perp_i \boldsymbol{S}_n )
		+
	\nonumber \\
		( \boldsymbol{S}_n \times \hat{\Delta}_i^\perp \boldsymbol{S}_n ) ( \boldsymbol{S}_n \times \hat{\Delta}_i^\perp \boldsymbol{S}_n )
		]
		\cdot \frac{\diff \boldsymbol{S}_n}{\diff t}
	\nonumber \\
		= S_n \Bigg(
			\alpha^2 
			+ \frac{\alpha \eta}{S_n^2} \sum_i | \hat{\Delta}^\perp_i \boldsymbol{S}_n |^2
		\Bigg)
		\frac{\diff \boldsymbol{S}_n}{\diff t}
	.
	\label{SD_part_4}
\end{gather}
Combining these four parts, we obtain
\begin{gather}
	\boldsymbol{S}_n \times \mathcal{D} \frac{\diff \boldsymbol{S}_n}{\diff t}
	=
	\alpha \boldsymbol{S}_n \times [ \boldsymbol{S}_n \times ( \boldsymbol{H}_n + \boldsymbol{H}_{\mathrm{R}} ) ]
	\nonumber \\
	- \alpha \beta S_n \boldsymbol{S}_n \times \boldsymbol{H}_{\mathrm{R}}
	\nonumber \\
	+ \frac{\eta}{S_n^2} \sum_i
		( \hat{\Delta}^\perp_i \boldsymbol{S}_n \cdot  \boldsymbol{H}_n )
		\boldsymbol{S}_{n} \times
		( \boldsymbol{S}_n \times \hat{\widetilde{\Delta}}_i \boldsymbol{S}_n )
	\nonumber \\
	-2 \alpha  \boldsymbol{S}_n \times ( \boldsymbol{S}_n \times \mathcal{J}^\mathrm{S}_n \boldsymbol{S}_n ) 
	\nonumber \\
	+	\alpha S_n \boldsymbol{S}_n \times \widetilde{\boldsymbol{H}}_n
		+ \alpha \beta \boldsymbol{S}_n \times ( \boldsymbol{S}_n \times \widetilde{\boldsymbol{H}}_n )
	\nonumber \\
		+ \frac{\eta}{S_n^3} \sum_i
			\boldsymbol{S}_n \cdot ( \hat{\widetilde{\Delta}}_i \boldsymbol{S}_n \times \widetilde{\boldsymbol{H}}_n )
			\boldsymbol{S}_n \times ( \boldsymbol{S}_n \times \hat{\widetilde{\Delta}}_i \boldsymbol{S}_n )
	\nonumber \\
	+ \alpha^2 S_n	
	\frac{\diff \boldsymbol{S}_n}{\diff t}
	.
	\label{SDdsdt}
\end{gather}
We have omitted the term involving $( \alpha \eta / S_n^2 ) \sum_i | \hat{\Delta}^\perp_i \boldsymbol{S}_n |^2$
coupled to the time derivative
since this term is at most comparable to $\alpha^2$\cite{bib:2058,bib:2347}, which is much smaller than unity,
and it would not affect significantly the entire time derivative.
Decomposing $\hat{\Delta}^\perp_i \boldsymbol{S}_n$ as Eq. (\ref{nderiv_2}) and 
$\boldsymbol{H}_n$ into components parallel and perpendicular to $\hat{\widetilde{\Delta}}_i \boldsymbol{S}_n$
as $\boldsymbol{H}_n = \boldsymbol{H}_{ni}^{\parallel} + \boldsymbol{H}_{ni}^\perp$,
we can write, using Eq. (\ref{hat_para_Delta}),
$
( \hat{\Delta}^\perp_i \boldsymbol{S}_n \cdot \boldsymbol{H}_n ) 
= ( \hat{\widetilde{\Delta}}_i \boldsymbol{S}_n \cdot \boldsymbol{H}_n ) 
- \Delta^\parallel_{i n} ( \boldsymbol{S}_n \cdot \boldsymbol{H}_{n i}^\parallel ) 
- \Delta^\parallel_{i n} ( \boldsymbol{S}_n \cdot \boldsymbol{H}_{n i}^\perp )
$.
By defining the two kinds of functionals as
$
\boldsymbol{A}_{0} [ \boldsymbol{S}; \boldsymbol{C}, a, b ] \equiv
a \boldsymbol{S} \times \boldsymbol{C}
+ b \boldsymbol{S} \times ( \boldsymbol{S} \times \boldsymbol{C} )
$
and
$
\boldsymbol{A}_{1} [ \boldsymbol{S}; \boldsymbol{C}, \boldsymbol{b} ] \equiv
( \boldsymbol{b} \cdot \boldsymbol{S} ) \boldsymbol{S} \times ( \boldsymbol{S} \times \boldsymbol{C} )/ S
$
, and substituting Eq. (\ref{SDdsdt}) back into the original equation, Eq. (\ref{LLGeq_1}),
we finally reach the following form:
\begin{gather}
	\frac{\diff \boldsymbol{S}}{\diff t} =
		\boldsymbol{A}^{(\mathrm{eff})} +
		\boldsymbol{A}^{(\mathrm{MCA})} +
		\boldsymbol{A}^{(\mathrm{SPC})} +
		\boldsymbol{A}^{(\mathrm{SOI})} 
	\nonumber \\
		+
		\boldsymbol{A}^{(\mathrm{ED} 0)} +
		\boldsymbol{A}^{(\mathrm{ED} \parallel )} +
		\boldsymbol{A}^{(\mathrm{ED} \perp )} 
	,
	\label{LLGeq_rewritten}
\end{gather}
where
\begin{gather}
	\boldsymbol{A}^{(\mathrm{eff})}
	= \boldsymbol{A}_{0} [ \boldsymbol{S}; \boldsymbol{H}, \kappa, -\kappa \alpha/S] 
	,
	\\
	\boldsymbol{A}^{(\mathrm{MCA})} 
	=	- 2 \kappa  \boldsymbol{S} \times \mathcal{J}^\mathrm{S} \boldsymbol{S}
		+ 2  \frac{\kappa \alpha}{S} \boldsymbol{S} \times ( \boldsymbol{S} \times \mathcal{J}^\mathrm{S} \boldsymbol{S} )
	,
	\label{eom_MCA}
	\\
	\boldsymbol{A}^{(\mathrm{SPC})} 
	= \boldsymbol{A}_{0} [ \boldsymbol{S}; \widetilde{\boldsymbol{H}}, \kappa (-\alpha + \beta), -\kappa (\alpha \beta + 1)/S ] 
	,
	\\
	\boldsymbol{A}^{(\mathrm{SOI})} 
	= \boldsymbol{A}_{0} [ \boldsymbol{S}; \boldsymbol{H}_\mathrm{R}, \kappa (1 + \alpha \beta), \kappa (-\alpha + \beta)/S ]
	,
	\\
	\boldsymbol{A}^{(\mathrm{ED} 0)}
	= \sum_i \boldsymbol{A}_0 [ \boldsymbol{S} ; \hat{\widetilde{\Delta}}_i \boldsymbol{S} , 0, -\kappa \eta  ( \hat{\widetilde{\Delta}}_i \boldsymbol{S} \cdot \boldsymbol{H} )/S^3 ]
	,
	\\
	\boldsymbol{A}^{(\mathrm{ED} \parallel )}
	= \sum_i \boldsymbol{A}_1 [ \boldsymbol{S} ; \hat{\widetilde{\Delta}}_i \boldsymbol{S} , \kappa \eta  \Delta_i^\parallel  \boldsymbol{H}^\parallel_i /S^2 ]
	,
	\\
	\boldsymbol{A}^{(\mathrm{ED} \perp )} 
	= \sum_i \boldsymbol{A}_1 \Bigg[ \boldsymbol{S} ; \hat{\widetilde{\Delta}}_i \boldsymbol{S} ,
		\frac{\kappa \eta}{S^2}
		\Bigg(
			\Delta_i^\parallel \boldsymbol{H}_i^\perp - \frac{ \hat{\widetilde{\Delta}}_i \boldsymbol{S} \times \widetilde{\boldsymbol{H}}}{S}
		\Bigg)
	\Bigg]
	,
\end{gather}
and $\kappa  \equiv 1/( 1 + \alpha^2 )$.
We have omitted the subscript $n$.
The equation of motion obtained here consists of 13 parts,
which are only of three kinds, $\boldsymbol{A}_0$, $\boldsymbol{A}_1$, and the magnetocrystalline anisotropy terms.
This is the first main result of the present work.
$\boldsymbol{A}_0$ describes an ordinary precessing and damped motion of a spin.
$\boldsymbol{A}_1$ describes an abnormal damped motion, which cannot be expressed as a special case of $\boldsymbol{A}_0$.
Since $\eta$ is no longer coupled to the time derivative,
the damping enhancement can be treated on the same footing as the ordinary precessing and damped motion.

\section{Exact solutions}

We have obtained the equation of motion for a generic case, Eq. (\ref{LLGeq_rewritten}),
however,
we could not hope that its exact solution for arbitrary parameters are obtained.
The equation should thus be solved numerically.
Ma and Dudarev\cite{bib:2260} adopted the STD method\cite{bib:2271} for the numerical integration of the equation of motion of spins.
A first-order differential equation of a dynamical vector variable $\boldsymbol{x}$,
\begin{gather}
	\frac{\diff \boldsymbol{x}}{\diff t} 
		= ( \hat{L}_1 + \hat{L}_2 ) \boldsymbol{x}
	,
\end{gather}
where $\hat{L}_1$ and $\hat{L}_2$ are operators acting on $\boldsymbol{x}$,
has a formal solution 
$\boldsymbol{x} (t + \Delta t) = \exp [ ( \hat{L}_1 + \hat{L}_2 ) \Delta t ] \boldsymbol{x} (t)$.
The STD method allows one to decompose the time evolution operator as
\begin{gather}
	e^{ ( \hat{L}_1 + \hat{L}_2 ) \Delta t}
	= e^{ \hat{L}_1 \Delta t/2 } e^{ \hat{L}_2 \Delta t } e^{ \hat{L}_1 \Delta t/2 }
		+ \mathcal{O}(\Delta t^3)
	,
\end{gather}
which means that if we divide a complex equation of motion into sufficiently small parts whose exact solutions are known,
the numerical errors of a simulation of the dynamics are restricted within those due to the STD method itself, as small as third-order with respect to the time step.
The advantages of the STD method over other integration methods for spin dynamics are explained in detail in literature.\cite{bib:2260}
In what follows,
the exact solutions of the three kinds of equations of motion derived above are provided.
While the exact solution for $\boldsymbol{A}_0$ has been reported,
those for $\boldsymbol{A}_1$ and the magnetocrystalline anisotropy term with general parameters are,
to my best knowledge,
first to be provided as the second main result of the present work.

\subsection{Ordinary precession and damping term}

Since the exact solution of the equation of motion governed only by an $\boldsymbol{A}_0$ term, 
\begin{gather}
	\frac{\diff \boldsymbol{S}}{\diff t}
	= a \boldsymbol{S} \times \boldsymbol{C}
		+ b \boldsymbol{S} \times ( \boldsymbol{S} \times \boldsymbol{C} )
	,
\end{gather}
has been provided in an earlier work\cite{bib:2260},
only its expression is written here.
With the parameters
$
\boldsymbol{S}_0 \equiv \boldsymbol{S} (t=0),
\xi \equiv aC,
\zeta \equiv bCS,
\chi \equiv ( \boldsymbol{S}_0 \cdot \boldsymbol{C} ) /(SC),
$
it is given by
\begin{gather}
	\boldsymbol{S} 
	= 
	\frac{1}{C [ 1 + e^{2 \zeta t} + \chi ( 1 - e^{2 \zeta t}) ]} \cdot
	\nonumber \\
		\cdot [ 2  C e^{\zeta t} \cos \xi t \boldsymbol{S}_0
		+ 2  e^{\zeta t} \sin \xi t (\boldsymbol{S}_0 \times \boldsymbol{C})
	\nonumber \\
		+ S \{ 1 - e^{2 \zeta t} + \chi (1 + e^{2 \zeta t} - 2e^{\zeta t} \cos \xi t ) \} \boldsymbol{C} ]
	.
\end{gather}

\subsection{Abnormal damping term}

Let us consider the equation of motion governed only by an $\boldsymbol{A}_1$ term,
\begin{gather}
	\frac{\diff \boldsymbol{S}}{\diff t}
	= \frac{ \boldsymbol{b} \cdot \boldsymbol{S} }{S}
		\boldsymbol{S} \times ( \boldsymbol{S} \times \boldsymbol{C} )
	.
\end{gather}
We first assume $\boldsymbol{C}$ to be along the $z$-axis: $\boldsymbol{C} = C \boldsymbol{e}_z$.
Expressing the equation of motion with the spherical coordinates $\theta$ and $\phi$ for $\boldsymbol{S}$,
we immediately obtain $\phi = \mathrm{const.} \equiv \phi_0$.
For obtaining the analytic expression of $\theta$ as a function of $t$,
we have to examine two cases according to whether $\boldsymbol{b}$ is parallel or perpendicular to $\boldsymbol{C}$.
Even when $\boldsymbol{b}$ is neither parallel nor perpendicular to $\boldsymbol{C}$,
decomposition of $\boldsymbol{b}$ into the two components allows us to use the STD method,
which is the reason for the separation of $\boldsymbol{A}^{(\mathrm{ED} \parallel )}$ and $\boldsymbol{A}^{(\mathrm{ED} \perp )}$.

\subsubsection{For $\boldsymbol{b}$ parallel to $\boldsymbol{C}$}

When $\boldsymbol{b}$ is parallel to $\boldsymbol{C}$, 
the equation $\theta$ must satisfy is
\begin{gather}
	\frac{\diff \theta}{\diff t}
		= S  b_z C \cos \theta  \sin \theta
	.
\end{gather}
By integrating both sides, we obtain
$\theta = \mathrm{arctan} [ \tan \theta_0 \exp (S b_z C t) ]$,
where $\theta_0 \equiv \theta (t = 0)$.
If $b_z C > 0$,
$\theta = \pi/2$ is a stable stationary point
and $\theta = 0$ and $\pi$ are unstable stationary points.
If $b_z C < 0$,
$\theta = \pi/2$ is an unstable stationary point
and $\theta = 0$ and $\pi$ are stable stationary points.
(See Fig. \ref{fig_damp}).

The solution for $\boldsymbol{C}$ in an arbitrary direction is obtained by rotating correctly the solution given above.
Its expression is
\begin{gather}
	\boldsymbol{S}
	= \frac{ S \chi ( e^{-\zeta t} - 1) \boldsymbol{C} + C \boldsymbol{S}_0}{ C \sqrt{ \chi^2 e^{-2 \zeta t} + 1 - \chi^2 }  } 
	,
\end{gather}
where $\zeta \equiv  S ( \boldsymbol{b} \cdot \boldsymbol{C} )$ and $\chi \equiv ( \boldsymbol{S}_0 \cdot \boldsymbol{C} )/(SC)$.

\subsubsection{For $\boldsymbol{b}$ perpendicular to $\boldsymbol{C}$}

\begin{figure}[htbp]
\begin{center}
\includegraphics[keepaspectratio,width=5cm]{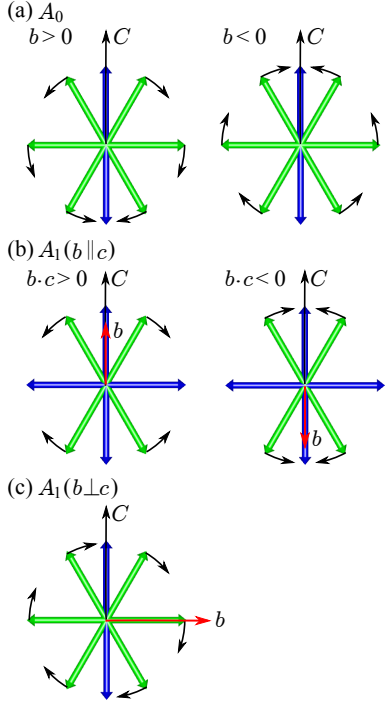}
\end{center}
\caption{
(Color online)
Directions of ordinary and abnormal damped motion of $\boldsymbol{S}$ for equations of motion governed by (a) $\boldsymbol{A}_0$ and (b), (c) $\boldsymbol{A}_1$.
Green (brighter) arrows represent spins at unstationary points,
while blue (darker) arrows represent stable or unstable stationary points.
In (a), precession is assumed to be absent ($a = 0$).
In (c), $\boldsymbol{b}$, $\boldsymbol{C}$ and $\boldsymbol{S}$ are assumed to lie on the same plane.
}
\label{fig_damp}
\end{figure}

When $\boldsymbol{b}$ is perpendicular to $\boldsymbol{C}$, 
the equation $\theta$ must satisfy is
\begin{gather}
	\frac{\diff \theta}{\diff t}
		= S  b_\perp C \sin^2 \theta
	,
\end{gather}
where $b_\perp \equiv  b_x \cos \phi_0 + b_y \sin \phi_0$.
By integrating both sides, we obtain
$\theta	= \mathrm{arctan} [ (\cot \theta_0 - S  b_\perp C t )^{-1} ]$.
If $b_\perp C > 0$,
$\theta = 0$ is an unstable stationary point
and $\theta = \pi$ is a stable stationary point.
If $b_\perp C < 0$,
$\theta = 0$ is a stable stationary point
and $\theta = \pi$ is an unstable stationary point.
It is noted that $\theta = \pi/2$ is not a stationary point.
(See Fig. \ref{fig_damp}).

The solution for $\boldsymbol{C}$ in an arbitrary direction is obtained by rotating correctly the solution given above.
Its expression is
\begin{gather}
	\boldsymbol{S}
	= \frac{ -S \zeta t \boldsymbol{C} + C \boldsymbol{S}_0 }{C \sqrt{ ( \chi - \zeta  t )^2 + 1 - \chi^2  }}
	,
\end{gather}
where $\zeta \equiv C (  \boldsymbol{b} \cdot \boldsymbol{S}_0 )$
and $\chi \equiv ( \boldsymbol{S}_0 \cdot \boldsymbol{C} )/(SC)$.

\subsection{Magnetocrystalline anisotropy term}

Let us consider the equation of motion governed only by a magnetocrystalline anisotropy term, Eq. (\ref{eom_MCA}),
\begin{gather}
	\frac{\diff \boldsymbol{S}}{\diff t} = 
		-2 \kappa \boldsymbol{S} \times \mathcal{J}^\mathrm{S} \boldsymbol{S}
		+2 \frac{\zeta}{S^2} \boldsymbol{S} \times ( \boldsymbol{S} \times \mathcal{J}^\mathrm{S} \boldsymbol{S} )
	,
	\label{eom_MCA2}
\end{gather}
where $\zeta \equiv \kappa \alpha S$.
This equation describes the dynamics of a spin feeling a torque coming from its own direction.
Since $\mathcal{J}^\mathrm{S}$ is a symmetric matrix, it can be diagonalized by a real orthogonal matrix $R$.
Let $\boldsymbol{v}_i (i = 1, 2, 3)$ be the eigenvector and $\lambda_i$ be the corresponding eigenvalue.
Then we can write
${}^\mathrm{t} R = ( \boldsymbol{v}_1, \boldsymbol{v}_2, \boldsymbol{v}_3 )$ and $\mathcal{J}^\mathrm{S} = {}^\mathrm{t} R D R$,
where $D \equiv \mathrm{diag} [\lambda_1, \lambda_2, \lambda_3 ]$.
${}^\mathrm{t} R$ is the transposition of $R$.
If the eigenvalues $\lambda_1$, $\lambda_2$, and $\lambda_3$ are different from each other,
we require that the eigenvalues be arranged according to the initial spin direction so that
\begin{gather}
	\begin{cases}
		\lambda_1 > \lambda_2 > \lambda_3  \, ( \boldsymbol{S}_0 \cdot \mathcal{J}^\mathrm{S} \boldsymbol{S}_0 - \lambda_2 S^2 < 0 ) \\
		\lambda_1 < \lambda_2 < \lambda_3  \, ( \boldsymbol{S}_0 \cdot \mathcal{J}^\mathrm{S} \boldsymbol{S}_0 - \lambda_2 S^2 > 0 ) \\
	\end{cases}
	\label{judge_lambda_2} 
\end{gather}
for later convenience for the $\zeta = 0$ case discussed below.

Eq. (\ref{eom_MCA2}) becomes via operation of $R$ on it from left as
\begin{gather}
	\frac{\diff \widetilde{S}_x}{\diff t}
		= - 2 \eta (\lambda_3 - \lambda_2) \widetilde{S}_y \widetilde{S}_z 
	\nonumber \\
			- 2 \frac{\zeta}{S^2}
		[ \lambda_1 S^2 - ( \lambda_1 \widetilde{S}_x^2 + \lambda_2 \widetilde{S}_y^2 + \lambda_3 \widetilde{S}_z^2 ) ] \widetilde{S}_x
	,
	\label{eom_MCA3x}
	\\
	\frac{\diff \widetilde{S}_y}{\diff t}
		= - 2 \eta (\lambda_1 - \lambda_3) \widetilde{S}_z \widetilde{S}_x 
	\nonumber \\
			- 2 \frac{\zeta}{S^2}
		[ \lambda_2 S^2 - ( \lambda_1 \widetilde{S}_x^2 + \lambda_2 \widetilde{S}_y^2 + \lambda_3 \widetilde{S}_z^2 ) ] \widetilde{S}_y
	,
	\label{eom_MCA3y}
	\\
	\frac{\diff \widetilde{S}_z}{\diff t}
		= - 2 \eta (\lambda_2 - \lambda_1) \widetilde{S}_x \widetilde{S}_y 
	\nonumber \\
			- 2 \frac{\zeta}{S^2}
		[ \lambda_3 S^2 - ( \lambda_1 \widetilde{S}_x^2 + \lambda_2 \widetilde{S}_y^2 + \lambda_3 \widetilde{S}_z^2 ) ] \widetilde{S}_z
	,
	\label{eom_MCA3z}
\end{gather}
where $\widetilde{\boldsymbol{S}} \equiv R \boldsymbol{S}$ and $\eta \equiv \kappa \det R$.
Whether this system of equations can be solved exactly depends on the eigenvalues.
If the three are equal, $\boldsymbol{S}$ does not change.
If two of them are equal and the other is different, the equations of motion, Eqs. (\ref{eom_MCA3x}) - (\ref{eom_MCA3z}), can be solved exactly.
If the three are different from each other, exact solutions cannot be obtained
and we resort to dividing the equation of motion into two parts, the precession and the damping terms,
whose exact solutions can be obtained separately.
Thanks to the STD method, this division only increase the number of time-development operators
and not acts as a source of numerical errors.
Thus the consideration for the following three cases suffices.

\subsubsection{For $\lambda_1 = \lambda_2$}

Here we consider a case in which two of the eigenvalues are equal.
We can assume that $\lambda_1 = \lambda_2 \equiv \lambda$ and $\det R = 1$ without loss of generality.
We use the spherical coordinates $\widetilde{\theta}$ and $\widetilde{\phi}$ of $\widetilde{\boldsymbol{S}}$.
The equation $\widetilde{\theta}$ must satisfy in this case is, from Eq. (\ref{eom_MCA3z}),
\begin{gather}
	\frac{\diff \widetilde{\theta}}{\diff t} =  2 \zeta  (\lambda_3 - \lambda ) \sin \widetilde{\theta} \cos \widetilde{\theta}
	.
	\label{dthdt_mca_eqeval}
\end{gather}
Integrating both sides, we obtain
\begin{gather}
	\widetilde{\theta} = \arctan [ \tan \widetilde{\theta}_0 \exp \{ 2 \zeta  (\lambda_3 - \lambda ) t \} ]
	.
	\label{theta_mca_eqeval}
\end{gather}
$\widetilde{\theta} = 0, \pi$, and the equator ($\widetilde{\theta} = \pi/2$) are the stationary points.
The stability of them are determined by the signature of $\lambda_3 - \lambda$.

The equation $\widetilde{\phi}$ must satisfy is, from Eqs. (\ref{eom_MCA3x}) and (\ref{eom_MCA3y}),
\begin{gather}
	\frac{\diff \widetilde{\phi}}{\diff t} 
	= 2 \eta (\lambda_3 - \lambda) S \cos \widetilde{\theta} 
	\label{phi_mca_eqeval}
	.
\end{gather}
If there is damping ($\alpha \ne 0$),
Eqs. (\ref{dthdt_mca_eqeval}) and (\ref{phi_mca_eqeval}) lead to
\begin{gather}
	\frac{\diff \widetilde{\phi}}{\diff \widetilde{\theta}} = \frac{1}{\alpha \sin \widetilde{\theta} }  
	,
\end{gather}
from which 
\begin{gather}
	\widetilde{\phi}
	= \widetilde{\phi}_0 + \frac{1}{\alpha}  \ln \frac{\tan (\widetilde{\theta}/2)}{\tan (\widetilde{\theta}_0/2)}
\end{gather}
is obtained.
On the other hand, if there is no damping ($\alpha = 0$), $\widetilde{\theta}$ is constant and
\begin{gather}
	\widetilde{\phi}
	= \widetilde{\phi}_0 + 2 \eta (\lambda_3 - \lambda) S t \cos \widetilde{\theta}_0
\end{gather}
is obtained.

\subsubsection{For $\alpha = 0$}

Here we consider a case in which only the precession terms are present in Eqs. (\ref{eom_MCA3x}) - (\ref{eom_MCA3z})
and the three eigenvalues are different from each other.
In this case the equation of motion is of the same form as that for a freely rotating classical rigid body, known as the Euler's top.
The equation governing the dynamics of the top is called the Euler's equation\cite{bib:Goldstein},
whose exact solution has been provided by Zon and J. Schofield.\cite{bib:2333}
According to them, we put the solution into the form,
\begin{gather}
	\widetilde{S}_{x} = \widetilde{S}_{xm} \mathrm{cn} ( \omega_p t + \varepsilon, k )
	,
	\label{mca_solution_cn}
	\\
	\widetilde{S}_{y} = \widetilde{S}_{ym} \mathrm{sn} ( \omega_p t + \varepsilon, k )
	, 
	\label{mca_solution_sn}
	\\
	\widetilde{S}_{z} = \widetilde{S}_{zm} \mathrm{dn} ( \omega_p t + \varepsilon, k )
	,
	\label{mca_solution_dn}
\end{gather}
where $\mathrm{cn}$, $\mathrm{sn}$, and $\mathrm{dn}$ are the Jacobi elliptic functions.\cite{bib:2333_21}
The order of the eigenvalues assumed above, Eq. (\ref{judge_lambda_2}),
is needed for this ansatz.
The exact solution of the equation of motion is obtained by setting
\begin{gather}
	\widetilde{S}_{xm} 
	= ( \mathrm{sgn}  \widetilde{S}_{x0} )
		\sqrt{ \widetilde{S}_{x0}^2 + \frac{\lambda_2 - \lambda_3}{\lambda_1 - \lambda_3} \widetilde{S}_{y0}^2 }
	,
	\\
	\widetilde{S}_{ym}
	= -( \mathrm{sgn} \widetilde{S}_{x0} )
		\sqrt{ \frac{\lambda_1 - \lambda_3}{\lambda_2 - \lambda_3} \widetilde{S}_{x0}^2 + \widetilde{S}_{y0}^2 }
	,
	\\
	\widetilde{S}_{zm}
	= ( \mathrm{sgn} \widetilde{S}_{z0} )
		\sqrt{ \frac{\lambda_2 - \lambda_1}{\lambda_3 - \lambda_1} \widetilde{S}_{y0}^2 + \widetilde{S}_{z0}^2 }
	,
	\\
	\omega_p = [ \mathrm{sgn} (\lambda_2 - \lambda_3) ] ( \mathrm{sgn} \widetilde{S}_{z0} )  2 \eta
		\cdot
	\nonumber \\
		\cdot 
		\sqrt{[ (\lambda_2 - \lambda_1) \widetilde{S}_{y0}^2 + (\lambda_3 - \lambda_1) \widetilde{S}_{z0}^2  ] (\lambda_3 - \lambda_2) }
	\\
	k^2	= \frac{(\lambda_1 - \lambda_3) \widetilde{S}_{x0}^2 + (\lambda_2 - \lambda_3) \widetilde{S}_{y0}^2 }{(\lambda_2 - \lambda_1) \widetilde{S}_{y0}^2 + (\lambda_3 - \lambda_1) \widetilde{S}_{z0}^2}
			\cdot \frac{\lambda_1 - \lambda_2}{\lambda_3 - \lambda_2}
\end{gather}
and $\varepsilon = F(\widetilde{S}_{y0}/\widetilde{S}_{ym} ; k)$,
where $F$ is the incomplete elliptic integral of the first kind,\cite{bib:2333_21}
\begin{gather}
	F(x ;  k) = \int_0^x \frac{\diff t}{\sqrt{(1-k^2t)(1-t^2)}}
	.
\end{gather}

The solution, Eqs. (\ref{mca_solution_cn}) - (\ref{mca_solution_dn}),
indicates that, if $\widetilde{\boldsymbol{S}}_0$ is along the $x$- or $y$- or $z$-axis,
the spin does not change.

\subsubsection{For $\eta = 0$}

Here we consider a case in which only the damping terms are present in Eqs. (\ref{eom_MCA3x}) - (\ref{eom_MCA3z}).

The equation $\widetilde{\phi}$ must satisfy is, from Eqs. (\ref{eom_MCA3x}) and (\ref{eom_MCA3y}),
\begin{gather}
	\frac{\diff \widetilde{\phi}}{\diff t} = 2 \zeta (\lambda_1 - \lambda_2) \cos \widetilde{\phi} \sin \widetilde{\phi}
	.
\end{gather}
Integrating both sides, we obtain
\begin{gather}
	\widetilde{\phi} = 
		\arctan [ \tan \widetilde{\phi}_0 \exp \{ 2 \zeta  (\lambda_1 - \lambda_2 ) t \} ] 
	.
\end{gather}
Eq. (\ref{eom_MCA3z}) thus reads
\begin{gather}
	\frac{\diff \widetilde{\theta}}{\diff t} 
	= \Bigg[ \lambda_3 - \frac{\lambda_1 + \lambda_2 e^{ 4 \zeta  (\lambda_1 - \lambda_2 ) t }  \tan^2 \widetilde{\phi}_0 }{1 + e^{ 4 \zeta  (\lambda_1 - \lambda_2 ) t }  \tan^2 \widetilde{\phi}_0 } \Bigg]
		\zeta \sin 2 \widetilde{\theta} 
	.
\end{gather}
Making use of an integral formula
\begin{gather}
	\int \frac{p + q e^{bt}}{1 + a e^{bt}} \diff t
	= \Bigg( -p + \frac{q}{a} \Bigg) \frac{1}{b} \ln | 1 + ae^{bt} | + pt + c
	,
\end{gather}
where $c$ is an integral constant, leads to
\begin{gather}
	\widetilde{\theta} 
	= \arctan \Bigg[ 
			\tan \widetilde{\theta}_0 e^{ 2 \zeta  (\lambda_3 - \lambda_1) t }
			\sqrt{\frac{ 1 + e^{4 \zeta  (\lambda_1 - \lambda_2 ) t}  \tan^2 \widetilde{\phi}_0 }{ 1 + \tan^2 \widetilde{\phi}_0 }}
		\Bigg]
	.
\end{gather}

\section{Conclusions}

In conclusion,
we rewrote the modified LLG equation applicable to both classical and quantum mechanical description of the dynamics
of spins by paying attention to their microscopic arrangement.
The rewritten equation of motion was demonstrated to consist of the three kinds of terms, each of which can be solved exactly.
Their solutions were explicitly provided, suitable for the STD method.
The abnormal damping term, which originates in the enhanced damping tensor, cannot be expressed as a special case of the ordinary damping term.
The present work will help one to develop a simulation code for spin dynamics applicable to systems in various and complex situations.

%\begin{acknowledgments} 
%\end{acknowledgments}

\end{document}